\documentstyle[12pt]{article}
\begin{document}

\tolerance=5000

\def\cL{{\cal L}}
\def\be{\begin{equation}}
\def\ee{\end{equation}}
\def\bea{\begin{eqnarray}}
\def\eea{\end{eqnarray}}
\def\tr{{\rm tr}\, }
\def\nn{\nonumber \\}
\def\gd{g^\dagger}
\def\e{{\rm e}}
\newcommand{\inv}[1]{\left[#1\right]_{\mbox{inv}}}
\newcommand\DS{D \hskip -3mm / \ }
\def\ds{\left( 1 + {M \over \lambda}
\e^{\lambda(\sigma^- - \sigma^+ )}
\right)}

\  \hfill 
\begin{minipage}{3.5cm}
NDA-FP-37 \\
%your preprint No. \\
July 1997 \\
%hep-th/yymmxxx \\
\end{minipage}

\ 

\vfill

\begin{center}

{\Large\bf Trace anomaly induced effective
action and 2d black holes for dilaton coupled 
supersymmetric theories}

\vfill

{\large\sc Shin'ichi NOJIRI}\footnote{
e-mail : nojiri@cc.nda.ac.jp}
and
{\large\sc Sergei D. ODINTSOV$^{\spadesuit}$}\footnote{
e-mail : odintsov@quantum.univalle.edu.co, \\
odintsov@kakuri2-pc.phys.sci.hiroshima-u.ac.jp}

\vfill

{\large\sl Department of Mathematics and Physics \\
National Defence Academy \\
Hashirimizu Yokosuka 239, JAPAN}

{\large\sl $\spadesuit$ 
Tomsk Pedagogical University \\
634041 Tomsk, RUSSIA \\
and \\
Dep.de Fisica \\
Universidad del Valle \\
AA25360, Cali, COLOMBIA \\
}

\vfill

{\bf ABSTRACT}

\end{center}

The action for 2d dilatonic supergravity with 
dilaton coupled matter and dilaton multiplets is 
constructed. Trace anomaly and anomaly induced effective 
action (in components as well as in supersymmetric form) 
for matter supermultiplet on bosonic background are found. 
The one-loop effective action and large-$N$ effective action 
for quantum dilatonic supergravity are also calculated.
Using induced effective action one can estimate the back-reaction 
of dilaton coupled matter to the classical black hole solutions 
of dilatonic supergravity. That is done on the example of 
supersymmetric CGHS model with dilaton coupled quantum matter
 where Hawking radiation which turns out to be zero is 
calculated. Similar 2d analysis maybe used to study spherically 
symmetric collapse for other models of 4d supergravity. 

\ 

\noindent
PACS: 04.60.-m, 04.70.Dy, 11.25.-w  

\newpage

\section{Introduction}

There are various motivations to study 2d gravitational theories 
and their black hole solutions (see \cite{12,RST,A,BPP,Rev} and 
references therein). First of all,
it is often easier to study 2d models at least as useful laboratories.
Second, starting from 4d Einstein-Maxwell-scalar theory and applying 
spherically symmetric reduction anzats \cite{8}, 
one is left with specific dilatonic gravity with dilaton coupled 
matter. Hence, in such case 2d gravity with dilaton coupled matter 
may describe the radial modes of the extremal dilatonic 
black holes in four dimensions \cite{9}. 
Similarly, dilaton coupled matter action appears in string inspired models.
Recently, dilaton dependent trace anomaly and induced effective action 
for dilaton coupled scalars have been studied in refs.\cite{6,4,5}.
That opens new possibilities in the study of black holes 
with back-reaction of quantum matter \cite{4}.

It could be extremely interesting to present supersymmetric generalization 
of the results \cite{6,4,5}. 
Above motivations are still valid in 
this case. Indeed, let us
consider the spherical reduction of 
$N=1$, $d=4$ supergravity theory to $d=2$ 
theory. In order to realize the spherical reduction, 
we assume that the metric has the following form:
\bea
\label{spmetric}
ds^2&=&\sum_{\mu\nu=0,1,2,3}g_{\mu\nu}dx^\mu dx^\nu \nn
&=&\sum_{m,n=t,r}g_{mn}(t,r)dx^mdx^n
+\e^{2\phi(t,r)}(d\theta^2 + \sin^2\theta d\varphi^2)\ .
\eea
The metric (\ref{spmetric}) 
can be realized by choosing the vierbein 
fields $e^a_\mu$ as follows
\bea
\label{spvier}
&& e^0_{\theta, \varphi}=e^3_{\theta, \varphi}=
e^1_{t,r}=e^2_{t,r}=0\ , \nn
&& e^1_\theta=\e^\phi\ ,\ \ 
e^2_\varphi=\sin\theta\e^\phi\ ,\ \ 
e^1_\varphi=e^2_\theta=0\ .
\eea
The expression (\ref{spvier}) is unique up to 
local Lorentz transformation.
The local supersymmetry transformation 
for the vierbein field with the papameter 
$\zeta$ and $\bar\zeta$ is given by, 
\be
\label{strnsf}
\delta e^a_\mu =i\left(\psi_\mu\sigma^a \bar\zeta
- \zeta\sigma^a \bar\psi_\mu\right)\ .
\ee
Here $\psi_\mu$ is Rarita-Schwinger field 
(gravitino) and we follow the standard notations 
of ref.\cite{WB}(see also \cite{GGRS}.
If we require that the metric has the form of  
(\ref{spmetric}) after the local supersymmetry 
transformation, i.e., 
\bea
\label{metric}
&&\delta g_{t\theta}=\delta g_{r\theta}
=\delta g_{t\varphi}=\delta g_{r\varphi}=
\delta g_{\theta\varphi}=0 \ , \nn
&& \delta g_{\varphi\varphi}
=\sin\theta \delta g_{\theta\theta}\ ,
\eea
we obtain, up to local Lorentz transformation,
\bea
\label{solpara}
&&\zeta_1 = \bar\zeta^1\ ,\ \ \zeta_2 = \bar\zeta^2\ ,\\
\label{solgrtn}
&& \psi_{\varphi 1}=\sin\theta\psi_{\theta 1}\ ,\ \ 
\bar\psi_\varphi^1=\sin\theta\bar\psi_\theta^1 \ , \nn
&& \psi_{\varphi 2}=-\sin\theta\psi_{\theta 2}\ ,\ \ 
\bar\psi_\varphi^2=-\sin\theta\bar\psi_\theta^2 \ , 
\nn
&& \bar\psi^1_\theta=-\psi_{\theta 1}\ ,\ \ 
\bar\psi^2_\theta=-\psi_{\theta 2}\ ,\nn
&&
\psi_{r1}-\bar\psi_r^1
=-2\e^{-\phi}\e^3_r\psi_{\theta 2}\ ,\ \ 
\psi_{t1}-\bar\psi_t^1
=-2\e^{-\phi}\e^3_t\psi_{\theta 2}
\ ,\nn
&& \psi_{r2}-\bar\psi_r^2
=-2\e^{-\phi}\e^0_r\psi_{\theta 1}\ ,\ \ 
\psi_{t2}-\bar\psi_t^2
=-2\e^{-\phi}\e^0_r\psi_{\theta 1}
\ .
\eea
Eq.(\ref{solpara}) tells that the local supersymmetry 
of the spherically reduced theory is $N=1$, 
which should be compared with the torus compactified 
case, where the supersymmetry becomes $N=2$.
Let the independent degrees of freedom of the 
Rarita-Schwinger fields be
\bea
\label{gravitino}
&& 2\psi_r^1\equiv\psi_{r1}+\bar\psi_r^1\ ,\ \ 
2\psi_r^2\equiv\psi_{r2}+\bar\psi_r^2\ ,\nn
&& 2\psi_t^1\equiv\psi_{t1}+\bar\psi_t^1\ ,\ \ 
2\psi_t^2\equiv\psi_{t2}+\bar\psi_t^2\ ,\\ 
\label{dilatino}
&&\chi_1\equiv\psi_{\theta 1}\ ,\ \ 
\chi_2\equiv\psi_{\theta 2}\ ,
\eea
then we can regard $\psi_{t,r}$ and $\chi$ 
to be the gravitino and the dilatino in the 
spherically reduced theory.

If we coupled the massless matter multiplet, 
the action of the spherically reduced theory 
is given by 
\bea
\label{spmaction}
S&=&-{1 \over 4}\int d^4 x d^\theta{\cal E}
\left(\bar{\cal D}\bar{\cal D}-8R \right)
\Phi_i^\dagger \Phi_i \nn
&\sim 4\pi&\int dr dt \sqrt g \e^{2\phi} 
\Bigl(-\partial_\mu A_i \partial^\mu A_i 
- {i \over 2}\left[\chi_i\sigma^\mu{\cal D}_\mu\bar\chi_i
+\bar\chi_i\bar\sigma^\mu{\cal D}_\mu\chi_i\right] \nn
&& +\cdots \Bigr)\ .
\eea
Here $\cdots$ denotes the terms containing the 
Rarita-Schwinger fields and dilatino. 
Note that in the spherically reduced theory, the 
dilaton filed $\phi$ couples with the matter fields.
Therefore if we like to investigate 2d dilaton 
supergravity as the spherically reduced theory, 
we need to couple the dilaton field to the matter 
multiplet.

The present work is organised as following. Next section 
is devoted to the construction of the Lagrangian for 
2d dilatonic supergravity with dilaton coupled 
matter and dilaton supermultiplets. Dilaton dependent 
trace anomaly and induced effective action (as well
as large-$N$ effective action for quantum dilatonic supergravity)
for matter multiplet are found in section 3. Black holes 
solutions and their properties are discussed for some specific models 
in section 4. Some conclusions are presented in final section.

\section{The action of 2d dilatonic supergravity with 
matter}

In the present section we are going to construct the action 
of 2d dilatonic supergravity with dilaton supermultiplet and 
with matter supermultiplet.
The final result is given in superfields as well as in components.

In order to construct the Lagrangian of two-dimensional dilatonic 
supergravity, we use the component formulation 
 of ref.\cite{HUY}.
The conventions and notations are given  as following:
\begin{itemize}
\item signature
\be
\label{signature}
\eta^{ab}=\delta^{ab}=\left(\begin{array}{cc}
1 & 0 \\
0 & 1 \end{array}\right)
\ee
\item gamma matrices
\bea
\label{gammam}
\gamma^a\gamma^b&=&
\delta^{ab}+i\epsilon^{ab}\gamma_5 \ , \nn
\sigma_{ab}&\equiv& {1 \over 4}[\gamma_a, \gamma_b]
={i \over 2}\epsilon_{ab}\gamma_5\ .
\eea
\item charge conjugation matrix $C$
\bea
\label{cc}
&&C\gamma_a C^{-1}=-\gamma_a^T \ , \nn
&& C=C^{-1}= -C^T \ , \nn
&& \bar\psi=-\psi^T C\ .
\eea
Here the index ${^T }$ means transverse.
\item Majorana spinor
\be
\psi=\psi^c\equiv C\bar\psi^T\ .
\ee
\item Levi-Civita tensor
\bea
\label{LC}
&&\epsilon^{12}=\epsilon_{12}=1\ ,
\ \ \epsilon^{ab}=-\epsilon^{ba}\ , 
\ \ \epsilon_{ab}=-\epsilon_{ba}\ , \nn
&& \epsilon^{\mu\nu}=e\, e_a^\mu
e_b^\nu \epsilon_{ab} 
\ \ \epsilon_{\mu\nu}=e^{-1}\, e^a_\mu e^b_\nu 
\epsilon^{ab} \ .
\eea
\end{itemize}
In this paper, all the scalar fields are real and 
all the spinor fields are Majorana spinors.

We introduce dilaton multiplet $\Phi=(\phi,\chi,F)$ 
and matter multiplet $\Sigma_i=(a_i,\chi_i,G_i)$,  
which has the conformal weight $\lambda=0$, and
the curvature multiplet $W$\footnote{
The definition of the scalar curvature $R$ is different 
from that of Ref.\cite{HUY} $R^{HUY}$ by sign
\[R=-R^{HUY}\ .\]}
\be
\label{W}
W=\left(S,\eta,-S^2-{1 \over 2}R-{1 \over 2}
\bar\psi^\mu\gamma^\nu\psi_{\mu\nu}
+{1 \over 4}\bar\psi^\mu\psi_\mu\right)\ .
\ee
Here $R$ is the scalar curvature and 
\bea
\label{eta}
\eta&=&-{1 \over 2}S\gamma^\mu\psi_\mu 
+{i \over 2}e^{-1}\epsilon^{\mu\nu}\gamma_5
\psi_{\mu\nu} \ , \\
\label{psimn}
\psi_{\mu\nu}&=&D_\mu\psi_\nu - D_\nu\psi_\mu \ , \nn
D_\mu\psi_\nu&=&\left(\partial_\mu 
-{1 \over 2}\omega_\mu\gamma_5\right)\psi_\nu \ , \nn
\omega_\mu&=&-ie^{-1}e_{a\mu}\epsilon^{\lambda\nu}
\partial_\lambda e^a_\nu
-{1 \over 2}\bar\psi_\mu\gamma_5\gamma^\lambda
\psi_\lambda\ .
\eea
The curvature multiplet has the conformal 
weight $\lambda=1$.

Then the general action of 2d dilatonic supergravity 
is given in terms of general functions of the 
dilaton $C(\phi)$, $Z(\phi)$, $f(\phi)$ and 
$V(\phi)$\footnote{
The multiplet containing $C(\phi)$, for example, 
is given by $(C(\phi), C'(\phi)\chi, C'(\phi)F
-{1 \over 2}C''(\phi)\bar\chi\chi)$. } as follows
\bea
\label{lag}
&&\cL=-\inv{C(\Phi)\otimes W} \nn
&& +{1 \over 2}
\inv{\Phi\otimes\Phi\otimes T_P(Z(\Phi))} 
-\inv{Z(\Phi)\otimes\Phi \otimes
T_P(\Phi)} \nn
&& +\sum_{i=1}^N
\left\{{1 \over 2}
\inv{\Sigma_i\otimes\Sigma_i\otimes T_P(f(\Phi))}
-\inv{f(\Phi)\otimes \Sigma_i\otimes T_P(\Sigma_i)}
\right\} \nn
&& +\inv{V(\Phi)}\ , \nn
&& \inv{C(\Phi)\otimes W} \nn 
&&= e\left[
C(\phi)\left(-S^2 - {1 \over 2}R-{1 \over 2}
\bar\psi^\mu\gamma^\nu\psi_{\mu\nu}\right) 
+ C'(\phi)(FS-\bar\chi\eta) \right. \nn
&& \left. -{1 \over 2}C''(\phi)\bar\chi\chi S 
+{1 \over 2}\bar\psi_\mu\gamma^\mu (\eta C(\phi)
+\chi S C'(\phi)) 
+{1 \over 2}C(\phi)S\bar\psi_\mu\sigma^{\mu\nu}
\psi_\nu \right] \ , \nn
&& \inv{\Phi\otimes\Phi\otimes T_P(Z(\Phi))} \nn
&&=e\left[ \left(Z'(\phi)F-{1 \over 2}Z''(\phi)\bar\chi\chi\right)
(2\phi F-\bar\chi\chi) 
+\phi^2\Box(Z(\phi))-2\phi\bar\chi\DS (Z'(\phi)
\chi) \right. \nn
&&+{1 \over 2}\bar\psi_\mu\gamma^\mu \left\{
2\chi \left(Z'(\phi)F
-{1 \over 2}Z''(\phi)\bar\chi\chi\right)\phi 
+\DS (Z'(\phi)\chi)\phi^2 \right\} \nn
&& \left. +{1 \over 2}\left(Z'(\phi)F
-{1 \over 2}Z''(\phi)\bar\chi\chi\right)
\phi^2\bar\psi_\mu\sigma^{\mu\nu}
\psi_\nu \right] \ , \nn
&& \inv{Z(\Phi)\otimes\Phi \otimes T_P(\Phi)} \nn
&&=e\Bigl[Z'(\phi)(F^2\phi-\bar\chi\chi F 
-\phi \bar\chi\DS\chi) \nn
&& -{1 \over 2}Z''(\phi)\bar\chi\chi\phi F 
+ Z(\phi)(\phi\Box\phi -\bar\chi\DS\chi + F^2) \nn
&& +{1 \over 2}\bar\psi_\mu\gamma^\mu \{
\DS\chi Z(\phi)\phi +\chi(Z(\phi)+Z'(\phi)\phi)F\}
+{1 \over 2}Z(\phi)\phi F 
\bar\psi_\mu\sigma^{\mu\nu}
\psi_\nu \Bigr] \ , \nn
&& \inv{f(\Phi)\otimes \Sigma_i\otimes T_P(\Sigma_i)} \nn
&&=e\Bigl[f'(\phi)(F a_i G_i-\bar\chi\xi_i G_i 
-a_i \bar\chi\DS\xi_i) \nn
&& -{1 \over 2}f''(\phi)\bar\chi\chi a_i G_i 
+ f(\phi)(a_i\Box a_i -\bar\xi_i\DS\xi_i + G_i^2) \nn
&& +{1 \over 2}\bar\psi_\mu\gamma^\mu \{
\DS\xi_i f(\phi)a_i +(\xi_i f(\phi)
+\chi f'(\phi)a_i)G_i\}
+{1 \over 2}Z(\phi)a_i G_i 
\bar\psi_\mu\sigma^{\mu\nu}
\psi_\nu \Bigr] \ , \nn
&&\inv{\Sigma_i\otimes\Sigma_i\otimes T_P(f(\Phi))} \nn
&&=e\left[\left(f'(\phi)F-{1 \over 2}f''(\phi)\bar\chi\chi\right)
(2a_i G_i -\bar\xi_i\xi_i) 
+a_i^2\Box(f(\phi))-2a_i\bar\xi_i\DS (f'(\phi)
\chi) \right. \nn
&&+{1 \over 2}\bar\psi_\mu\gamma^\mu \left\{
2\xi_i (f'(\phi)F
-{1 \over 2}f''(\phi)\bar\chi\chi)\a_i 
+\DS (f'(\phi)\chi)a_i^2 \right\} \nn
&& \left. +{1 \over 2}\left(f'(\phi)F
-{1 \over 2}f''(\phi)\bar\chi\chi\right)
a_i^2\bar\psi_\mu\sigma^{\mu\nu}
\psi_\nu \right]\ , \nn
&& \inv{V(\Phi)} \nn
&& = e\left[ V'(\phi)F 
- {1 \over 2}V''(\phi)\bar\chi\chi
+{1 \over 2}\bar\psi_\mu\gamma^\mu \chi V'(\phi) \right. \nn
&& \left. + \left({1 \over 2}\bar\psi_\mu\sigma^{\mu\nu}
\psi_\nu + S\right)V(\phi)\right]\ .
\eea
The covariant derivatives for the multiplet 
$Z=(\varphi, \zeta, H)$ with $\lambda=0$ are 
defined as
\bea
\label{covdrv}
D_\mu \varphi&=&\partial_\mu \varphi 
-{1 \over 2}\bar\psi_\mu\zeta \ , \nn
D_\mu \zeta &=& \left( \partial_\mu + 
{1 \over 2}\omega_\mu \gamma_5\right)\zeta
-{1 \over 2}D_\nu \varphi \gamma^\nu \psi_\mu 
- {1 \over 2}H\psi_\mu \ , \\
\Box \varphi &=& e^{-1}\left\{
\partial_\nu(e g^{\mu\nu}D_\mu \varphi) 
+ {i \over 4}\bar\zeta\gamma_5\psi_{\mu\nu}
\epsilon^{\mu\nu}
-{1 \over 2}\bar\psi^\mu D_\mu\zeta
-{1 \over 2}\bar\psi^\mu\gamma^\nu\psi_\nu 
D_\mu \varphi\right\}\ . \nonumber
\eea
$T_P(Z)$ is called the kinetic multiplet for 
the multiplet $Z=(\varphi, \zeta, H)$ and 
when the multiplet $Z$ has the comformal weight 
$\lambda=0$, $T_P(Z)$ has the following form
\be
\label{kin}
T_P(Z)=(H, \DS\zeta, \Box\varphi)\ .
\ee
The kinetic multiplet 
$T_P(Z)$ has conformal weight $\lambda=1$.
The product of two multiplets 
$Z_k=(\varphi_k, \zeta_k, H_k)$ $(k=1,2)$ 
with the conformal weight $\lambda_k$ is defined by 
\be
\label{product}
Z_1\otimes Z_2 =(\varphi_1\varphi_2, 
\varphi_1\zeta_2 + \varphi_2\zeta_1, 
\varphi_1 H_2 + \varphi_2 H_1 - \bar\zeta_1\zeta_2)\ .
\ee
The invariant Lagrangian $\inv{Z}$ for multiplet 
$Z$ is defined by
\be
\label{inv}
\inv{Z}=e\left[F+{1 \over 2}\bar\psi_\mu\gamma^\mu \zeta
+{1 \over 2}\varphi\bar\psi_\mu\sigma^{\mu\nu}\psi_\nu 
+S\varphi\right]\ .
\ee

In superconformal gauge
\be
\label{scgauge}
e^a_\mu =\e^\rho \delta^a_\mu\ (e=\e^\rho)\ ,
\ \ \psi_\mu=\gamma_\mu \psi\ 
(\bar\psi_\mu=-\bar\psi\gamma_\mu)\ ,
\ee
we find
\bea
\label{formula}
\omega_\mu&=&-i\epsilon^\lambda_{\ \mu}
\partial_\lambda\rho \ , \nn
eR&=&-2\partial_\mu\partial^\mu\rho \ , \nn
\epsilon^{\mu\nu}\psi_{\mu\nu}&=&
-2ie\gamma_5\gamma^\mu
\left(\partial_\mu - {1 \over 2}\partial_\mu\rho
\right)\psi \ , \nn
\eta&=& -S\psi + \gamma^\mu
\left(\partial_\mu - {1 \over 2}\partial_\mu\rho
\right)\psi \ , \nn
\bar\psi^\mu\gamma^\nu\psi_{\mu\nu}&=&
-2\bar\psi\left(\partial_\mu - {1 \over 2}\partial_\mu\rho
\right)\psi \ ,\nn
\bar\psi_\mu\sigma^{\mu\nu}\psi_\nu&=&-\bar\psi\psi\ .
\eea

Hence, we constructed the classical action for 2d 
dilatonic supergravity with dilaton and matter supermultiplets.
%Integation of the auxilliary fields $S$, $F$ and $G_I$ 
%in (\ref{lag}) 
%does not give any contribution since there is no 
%term linear in the auxilliary fields.

\section{Effective action in large-$N$ 
approach on bosonic background}

Our purpose in this section will be the study of trace anomaly 
and effective action in large-$N$ 
approximation for the 2d dilatonic supergravity 
discussed in previous section. 
We consider only bosonic background below as it will 
be sufficient 
for our purposes (study of black hole type solutions).

 On the bosonic  
background where dilatino $\chi$ and 
the Rarita-Schwinger fields vanish, one can 
show that 
the gravity and dilaton part of the Lagrangian have
the following form:
\bea
\label{gdlag}
&&\inv{C(\Phi)\otimes W}  \nn
&&= e\left[-C(\phi)\left(S^2 +{1 \over 2}R\right)
-C'(\phi)FS\right] \ , \nn
&& \inv{\Phi\otimes\Phi\otimes T_P(Z(\Phi))} \nn
&&=e\left[\phi^2\tilde\Box(Z(\phi)) 
+ 2Z'(\phi)\phi F^2\right] \ , \nn
&& \inv{Z(\Phi)\otimes\Phi \otimes T_P(\Phi)} \nn
&&=e\left[Z(\phi)\phi\tilde\Box\phi 
+ Z'(\phi)\phi F^2 + Z(\phi)F^2 \right] \ , \nn
&& \inv{V(\Phi)} \nn
&& = e\left[ V'(\phi)F  + SV(\phi)\right]\ .
\eea
For matter part we obtain
\bea
\label{bg}
&&\inv{f(\Phi)\otimes \Sigma_i\otimes T_P(\Sigma_i)} \nn
&&=e\left[ f(\phi)(a_i\tilde\Box a_i 
-\bar\xi_i\tilde\DS\xi_i ) + f'(\phi)Fa_iG_i 
+ f(\phi)G_i^2\right] \ , \nn
&& \inv{\Sigma_i\otimes\Sigma_i\otimes T_P(f(\Phi))} \nn
&&=e\left[a_i^2\tilde\Box(f(\phi))
+ 2f'(\phi)Fa_iG_i\right]\ .
\eea
Here 
the covariant derivatives for the multiplet 
$(\varphi, \zeta, H)$ with $\lambda=0$ are 
reduced to
\bea
\label{covdrv2}
\tilde D_\mu \varphi&=&\partial_\mu \varphi \ , \nn
\tilde D_\mu \zeta &=& \left( \partial_\mu + 
{1 \over 2}\omega_\mu \gamma_5\right)\zeta \ , \nn
\tilde\Box \varphi &=& e^{-1}
\partial_\nu(e g^{\mu\nu}\partial_\mu \varphi) \ .
\eea
 Using equations of motion with respect to 
the auxilliary fields $S$, $F$, $G_i$, 
 on the bosonic background one can show that 
\bea
\label{bgaf}
S&=&{C'(\phi)V'(\phi) - 2V(\phi)Z(\phi) 
\over {C'}^2(\phi) + 4C(\phi) Z (\phi)} \ , \nn
F&=&{C'(\phi) V(\phi)+ 2 C(\phi)V'(\phi)
\over {C'}^2(\phi) + 4C(\phi) Z (\phi)} \ , \nn
G_i&=&0\ .
\eea
 We will be interested in the supersymmetric extension \cite{NO} of 
the CGHS model \cite{12} as in specifical example for study 
of black holes and Hawking radiation. For such a model 
\be
\label{CGHS}
C(\phi)=2\e^{-2\phi}\ ,\ \ Z(\phi)=4\e^{-2\phi}\ ,
\ \ V(\phi)=4\e^{-2\phi}\ ,
\ee
we find
\be
\label{CGHSaf}
S=0\ ,\ \ F=-\lambda\ ,\ \ G_i=0\ .
\ee

Using (\ref{bg}), we can show that  
\bea
\label{lag2}
&& \sum_{i=1}^N
\left\{{1 \over 2}
\inv{\Sigma_i\otimes\Sigma_i\otimes T_P(f(\Phi))}
-\inv{f(\Phi)\otimes \Sigma_i\otimes T_P(\Sigma_i)}\right\} \nn
&&=ef(\phi)\sum_{i=1}^N
(g^{\mu\nu}\partial_\mu a_i \partial_\nu a_i
+\bar\xi_i\gamma^\mu\partial_\mu\xi_i 
-f(\phi)G_i^2) \nn
&& + \ \mbox{total divergence terms}\ .
\eea
Here we have used the fact that
\be
\label{maj}
\bar\xi_i \gamma_5 \xi=0
\ee
for the Majorana spinor $\xi_i$.

Let us start now the investigation of effective action 
in above theory. It is clearly seen that theory 
(\ref{lag2}) is conformally invaiant on the 
gravitational background under discussion.
Then using standard methods, we can prove that theory 
with matter multiplet $\Sigma_i$ is superconformally 
invariant theory. 
First of all, one can find trace anomaly $T$ 
for the theory (\ref{lag2}) on gravitational 
background using the following relation
\be
\label{gamma}
\Gamma_{div}={1 \over n-2}\int d^2x \sqrt g b_2\ , \ 
\ \ \ T=b_2
\ee
where $b_2$ is $b_2$ coefficient of 
Schwinger-De Witt expansion and $\Gamma_{div}$ 
is one-loop effective action.
The calculation of $\Gamma_{div}$ (\ref{gamma}) for 
quantum theory with Lagrangian (\ref{lag2}) has 
been done some time ago in ref.\cite{6}.
Using results of this work, we find 
\bea
\label{tracean}
T&=&{1 \over 24\pi}\left\{{3 \over 2}NR 
- 3N\left({f'' \over f} - {{f'}^2 \over 2f^2}\right)
(\nabla^\lambda\phi)(\nabla_\lambda\phi) \right. \nn
&& \left. - 3N{f' \over f}\Delta\phi\right\}
\eea
It is remarkable that Majorana spinors do not give 
the contribution to the dilaton dependent terms in 
trace anomaly as it was shown in \cite{6}. 
They only alter the coefficient of curvature 
term in $T$ (\ref{tracean}). 
Hence, except the coefficient of curvature term in 
$T$ (\ref{tracean}), the trace anomaly (\ref{tracean}) 
coincides with the correspondent expression for dilaton
coupled scalar \cite{4}.
Note also that for particular case 
$f(\phi)=\e^{-2\phi}$ the trace anomaly for 
dilaton coupled scalar has been recently calculated in 
refs.\cite{5}.

Making now the conformal transformation of the 
metric $g_{\mu\nu}\rightarrow \e^{2\sigma}g_{\mu\nu}$ 
in the trace anomaly, and using relation:
\be
\label{TW}
T={1 \over \sqrt g}{\delta \over \delta \sigma}
W[\sigma]
\ee
one can find anomaly induced action $W[\sigma]$.
In the covariant, non-local form it may be found as 
following:
\be
\label{qc}
W=-{1 \over 2}\int d^2x \sqrt{g} \Bigl[ 
{N \over 32\pi}R{1 \over \Delta}R 
-{N \over 16\pi}{{f'}^2 \over f^2}
\nabla^\lambda \phi
\nabla_\lambda \phi {1 \over \Delta}R 
-{N \over 8\pi}\ln f R \Bigr]\ .
\ee
Hence, we got the anomaly induced effective action 
for dilaton coupled matter multiplet in the 
external dilaton-gravitational background. 
We should note that the same action $W$ (\ref{qc}) 
gives the one-loop large-$N$ effective action in the 
quantum theory of supergravity with matter (\ref{lag}) 
(i.e., when all fields are quantized).
%%%

We can now rewrite $W$ in a supersymmetric way.
In order to write down the effective action expressing 
the trace anomaly, we need the supersymmetric extention 
of ${ 1 \over \Delta}R$. The extension is given by 
using the inverse kinetic multiplet in \cite{MNSN}, 
or equivalently by introducing two auxiliary field 
$\Theta=(t,\theta,T)$ and $\Upsilon=(u,\upsilon,U)$.
We can now construct the following action
\be
\label{TU}
\inv{\Theta\otimes (T_P(\Upsilon)-W)}\ .
\ee
The $\Theta$-equation of motion tells that, in the 
superconformal gauge (\ref{scgauge})
\be
\label{rp}
u\sim \rho \sim -{1 \over 2\Delta}R
\ , \ \ \ \upsilon\sim \psi\ .
\ee
Then we find 
\bea
\label{actions}
&& \sqrt g R{1 \over \Delta}R\sim 
4\inv{W\otimes \Upsilon} \nn
&& \sqrt g {{f'}^2(\phi) \over f^2(\phi)}
\nabla_\lambda\phi \nabla^\lambda \phi
{1 \over \Delta}R \nn
&& \sim -\inv{\Phi\otimes\Phi 
\otimes T_P\left({{f'}^2(\Phi) \over f^2(\Phi)}
\otimes \Upsilon \right)}\nn
&& \hskip 1cm + 2\inv{{{f'}^2(\Phi) \over f^2(\Phi)}
\otimes \Upsilon \otimes \Phi \otimes T_P(\Phi)} \ , \nn
&& \sqrt g \ln f(\phi) \, R\sim  
2\inv{\ln f(\Phi)\otimes W}\ .
\eea
In components,
\bea
\label{comTU}
&&\inv{\Theta\otimes (T_P(\Upsilon)-W)} \nn
&& =e\left[ t\left( \Box u + {1 \over 2}R 
+ {1 \over 2}\bar\psi^\mu\gamma^\nu\psi_{\mu\nu}
-S\bar\psi^\mu\psi_\mu\right) 
+T(U-S)-\bar\theta(\DS\upsilon - \eta) \right. \nn
&& \ \left. +{1 \over 2}\bar\psi_\mu\gamma^\mu\{
(\DS\upsilon - \eta)t+\theta(U-S)\} 
+{1 \over 2}t(U-S)\bar\psi_\mu\sigma^{\mu\nu}
\psi_\nu\right] \ , \\
\label{comWU}
&&4\inv{W\otimes \Upsilon} \nn
&&=4e\left[u\left( -{1 \over 2}R 
-{1 \over 2}\bar\psi^\mu\gamma^\nu\psi_{\mu\nu}
+S\bar\psi^\mu\psi_\mu\right) 
 +US-\bar\upsilon\eta \right. \nn
&& \ \left. +{1 \over 2}\bar\psi_\mu\gamma^\mu\{
\eta u +\upsilon S\} 
+{1 \over 2} uS\bar\psi_\mu\sigma^{\mu\nu}\psi_\nu
\right] \ , \\
\label{comPPTU}
&& \inv{\Phi\otimes\Phi 
\otimes T_P\left({{f'}^2(\Phi) \over f^2(\Phi)}
\otimes \Upsilon \right)} \nn
&& =e\Bigl[ \phi^2 \Box (u{h'}^2(\phi)) \nn
&& \ +(2\phi F - \bar\chi\chi )\Bigl\{
{h'}^2(\phi)U \nn
&& \ \ +u\{2h'(\phi)h''(\phi)F - 
({h''}^2(\phi)+h'(\phi)h'''(\phi))\bar\chi\chi\}
\Bigr\} \nn
&& \ -2\phi\bar\chi\DS({h'}^2(\phi)\upsilon 
+ 2u h'(\phi)h''(\phi)\chi) \nn
&& \ +{1 \over 2}\bar\psi_\mu\gamma^\mu\Bigl(
\{\DS({h'}^2(\phi)\upsilon + 2u h'(\phi)h''(\phi)\chi) 
\}\phi^2 \nn
&& \ +2\chi\phi\Bigl\{
{h'}^2(\phi)U+u\{2h'(\phi)h''(\phi)F - 
({h''}^2(\phi)+h'(\phi)h'''(\phi))\bar\chi\chi\}
\Bigr\}\Bigr) \nn
&& \ +{1 \over 2}\phi^2\Bigl\{
{h'}^2(\phi)U \nn
&& \ \ +u\{2h'(\phi)h''(\phi)F - 
({h''}^2(\phi)+h'(\phi)h'''(\phi))\bar\chi\chi\}
\Bigr\}\bar\psi_\mu\sigma^{\mu\nu}
\psi_\nu\Bigr] \ , \\
\label{comPPU}
&& \inv{{{f'}^2(\Phi) \over f^2(\Phi)}
\otimes \Upsilon \otimes \Phi \otimes T_P(\Phi)} \nn
&& = e\Bigl[u{h'}^2(\phi)\phi\Box\phi \nn
&& \ +F\Bigl\{\phi({h'}^2(\phi)U+u\{2h'(\phi)h''(\phi)F 
\nn
&& \ \ \ \ 
-({h''}^2(\phi)+h'(\phi)h'''(\phi))\bar\chi\chi\}
-2h'(\phi)h''(\phi)\bar\chi\upsilon) \nn
&& \ \ +u{h'}^2(\phi)F-\bar\chi(\upsilon {h'}^2(\phi)
+2\chi uh'(\phi)h''(\phi))\Bigr\} \nn
&& \ -(u{h'}^2(\phi)\bar\chi + \phi {h'}^2(\phi)
\bar\upsilon 
+2u\phi h'(\phi)h''(\phi)\bar\chi)\DS \chi \nn
&& \ +{1 \over 2}\bar\psi_\mu\gamma^\mu\left\{
\chi u{h'}^2(\phi)\phi + (\chi u{h'}^2(\phi) 
+\upsilon \phi {h'}^2(\phi) 
+2\chi u\phi h'(\phi)h''(\phi))F\right\} \nn
&& \ +{1 \over 2}u{h'}^2(\phi)\phi F
\bar\psi_\mu\sigma^{\mu\nu}
\psi_\nu\Bigr] \ , \\
\label{comPR}
&& \inv{\ln f(\Phi)\otimes W} \nn
&& =e\left[
h(\phi)\left(-S^2 + {1 \over 2}R-{1 \over 2}
\bar\psi^\mu\gamma^\nu\psi_{\mu\nu}\right) 
+ h'(\phi)(FS-\bar\chi\eta)
-{1 \over 2}h''(\phi)\bar\chi\chi S \right. \nn
&& \ \left. +{1 \over 2}\bar\psi_\mu\gamma^\mu (\eta h(\phi)
+\chi S h'(\phi)) 
+{1 \over 2}h(\phi)S\bar\psi_\mu\sigma^{\mu\nu}
\psi_\nu \right] \ .\nn
\eea
Here
\be
\label{h}
h(\phi)\equiv\ln f(\phi) \ . 
%\nn s(\phi)&\equiv&{{f'}^2(\Phi) \over f(\Phi)} - f''(\Phi)
\ee

That finishes the construction of large-$N$ supersymmetric 
anomaly induced effective 
action for 2d dilatonic supergravity with matter.

At the end of the present section, we will find the 
one-loop effective action (its divergent part) for the whole quantum theory (\ref{gdlag}), (\ref{bg}) on the 
bosonic background under discussion. Using 
Eqs. (\ref{gdlag}), (\ref{bg}), one can write the 
complete Lagrangian as following:
\bea
\label{clag}
e^{-1}L&=-& \Bigl( \tilde V + \tilde C R 
+ {1 \over 2}\tilde Z (\nabla_\mu\phi)(\nabla^\mu \phi) 
\nn
&& - f(\phi)\sum_{i=1}^N\left((\nabla_\mu a_i)
(\nabla^\mu a_i) + \bar\xi_i \gamma^\mu \partial_\mu 
\xi_i \right)
\eea
where
\bea
\label{tilpot}
-\tilde V &=& -CS^2 - C'FS + 2Z'\phi F^2 + Z'\phi F^2 
+ZF^2 \nn
&& + V'F + SV\ , \nn
\tilde C &=&{C \over 2}\ , \nn
2\tilde Z &=& 3\phi Z' + Z
\eea
where auxilliary fields equations of motion which 
lead to (\ref{bgaf}) should be used.

The calculation of the one-loop effective action 
for the theory (\ref{clag}) has been given in 
ref.\cite{6} in the harmonic gauge with the 
following result
\bea
\label{Gamma2}
\Gamma_{div}&=&-{1 \over 4\pi(n-2)}
\int d^2x \sqrt g \left\{ {48-3N \over 12}R 
+ {2 \over \tilde C}\tilde V 
+ {2 \over \tilde C'}\tilde V' \right. \nn
&& + \left({\tilde C'' \over \tilde C}
- {3\tilde {C'}^2 \over \tilde C^2} 
- {\tilde C'' \tilde Z \over \tilde {C'}^2}
- {N{f'}^2 \over 4 f^2} 
+ {N f'' \over 2f}\right)(\nabla^\lambda\phi)
(\nabla_\lambda\phi) \\
&& + \left. \left({\tilde C' \over \tilde C} 
- {\tilde Z \over \tilde C'} 
+ { Nf' \over 2f}\right) \Delta \phi 
-\left({3f\tilde Z \over 4\tilde {C'}^2} 
+{3f \over 4\tilde C} -{f' \over \tilde C'}\right)
\sum_{i=1}^N\bar\xi_i\gamma^\mu\partial_\mu\xi_i
\right\} .\nonumber
\eea
Thus, we found the one-loop effective action for
dilatonic supergravity with matter on bosonic 
background.
Of course, the contribution of fermionic 
superpartners is missing there.
However, Eq.(\ref{Gamma2}) gives also the divergent 
one-loop effective action in large-$N$ approximation
(one should keep only terms with multiplier $N$).
This effective action may be used also for construction 
of renormalization group improved effective 
Lagrangians and study of their properties like 
BH solutions in ref.\cite{NO2}.

%%%%%%%%%%%%%%%%%
\section{Black holes in supersymmetric 
extension of CGHS model 
with matter back reaction}

In the present section we discuss the particular 2d 
dilatonic supergravity model which represents 
the supersymmetric extension of CGHS model. Note 
that as a matter we use dilaton coupled matter supermultiplet. 
We would like to estimate back-reaction of such matter 
supermultiplet to black holes and Hawking radiation 
working in large-$N$ approximation.
Since we are interesting in the vacuum (black hole) 
solution, we consider the background where 
matter fields, the Rarita-Schwinger field and 
dilatino vanish.

In the superconformal gauge 
the equations of motion can be obtained by the variation 
over $g^{\pm\pm}$, $g^{\pm\mp}$ and $\phi$  
\bea
\label{eqnpp}
0&=&T_{\pm\pm} \nn
&=&\e^{-2\phi}\left(4\partial_\pm \rho
\partial_\pm\phi - 2 \left(\partial_\pm\phi\right)^2 
\right) \nn
&& +{N \over 8}\left( \partial_\pm^2 \rho 
- \partial_\pm\rho \partial_\pm\rho \right) \nn
&& +{N \over 8} \left\{
\left( 
\partial_\pm h(\phi) \partial_\pm h(\phi) \right)
\rho+{1 \over 2}{\partial_\pm \over \partial_\mp}
\left( \partial_\pm h(\phi) 
\partial_\mp h(\phi) \right)\right\} \nn
&& +{N \over 8}\left\{ 
-2 \partial_\pm \rho \partial_\pm h(\phi) 
+\partial_\pm^2 h(\phi) \right\} + t^\pm(x^\pm) \nn
&& + {N \over 64}{\partial_\pm \over \partial_\mp}
\left( h'(\phi)^2F^2 \right) \ , \\
\label{req}
0&=&T_{\pm\mp} \nn
&=&\e^{-2\phi}\left(2\partial_+
\partial_- \phi -4 \partial_+\phi\partial_-\phi 
- \lambda^2 \e^{2\rho}\right) \nn
&& -{N \over 8}\partial_+\partial_- \rho
-{N \over 16}\partial_+ h(\phi) 
\partial_- h(\phi) 
-{N \over 8}\partial_+\partial_-h(\phi) \nn
&& -{N \over 64}h'(\phi)^2F^2 
+\left({N \over 16}US 
+ {N \over 2}(-h(\phi)S^2 + h'(\phi)FS)\right)\e^{2\rho} \ , \\
\label{eqtp}
0&=& \e^{-2\phi}\left(-4\partial_+
\partial_- \phi +4 \partial_+\phi\partial_-\phi 
+2\partial_+ \partial_- \rho
+ \lambda^2 \e^{2\rho}\right) \nn
&& -{Nf' \over f}\left\{
{1 \over 16}\partial_+(\rho \partial_-h(\phi))
+{1 \over 16}\partial_-(\rho \partial_+h(\phi))
-{1 \over 8}\partial_+\partial_-\rho \right\} \ .
\eea
Here $t^\pm(x^\pm)$ is a function which is determined by 
the boundary condition.
Note that there is, in general, a contribution from 
the auxilliary fields to $T_{\pm\mp}$ besides the 
contribution from the trace anomaly.

In large-$N$ limit, where 
classical part can be ignored, field equations 
become simpler  
\bea
\label{eqnpp2}
0&=&{1 \over N}T_{\pm\pm} \nn
&=&{1 \over 8}\left( \partial_\pm^2 \rho 
- \partial_\pm\rho \partial_\pm\rho \right) \nn
&& +{1 \over 8} \left\{
\left( 
\partial_\pm h(\phi) \partial_\pm h(\phi) \right)
\rho+{1 \over 2}{\partial_\pm \over \partial_\mp}
\left( \partial_\pm h(\phi) 
\partial_\mp h(\phi) \right)\right\} \nn
&& +{1 \over 8}\left\{ 
-2 \partial_\pm \rho \partial_\pm h(\phi) 
+\partial_\pm^2 h(\phi) \right\} + t^\pm(x^\pm) \ , \\
\label{req2}
0&=&{1 \over N}T_{\pm\mp} \nn
&=& -{1 \over 8}\partial_+\partial_- \rho
-{1 \over 16}\partial_+ h(\phi) 
\partial_- \tilde \phi 
-{1 \over 8}\partial_+\partial_-h(\phi) \\
\label{eqtp2}
0&=&{1 \over 16}\partial_+(\rho \partial_-h(\phi))
+{1 \over 16}\partial_-(\rho \partial_+h(\phi))
-{1 \over 8}\partial_+\partial_-\rho\ . 
\eea
Here we used the $\Theta$-equation and the equations 
for the auxilliary fields $S$ and $F$, i.e.,
\be
\label{eqs}
U=S\ ,\ \ u=\rho=-{1 \over 2\Delta}R\ ,\ \ S=F=0\ .
\ee
The function $t^\pm(x^\pm)$ in (\ref{eqnpp2}) 
can be absorbed into 
the choice of the coordinate and we can choose
\be
\label{tpm}
t^\pm(x^\pm)=0\ .
\ee
Combining (\ref{eqnpp2}) and (\ref{req2}), 
we obtain
\be
\label{eqcm}
-{1 \over 2}(\partial_\pm \rho)^2 
+{1 \over 2}\rho(\partial_\pm h(\phi))^2
-\partial_\pm \rho \partial_\pm h(\phi) =0
\ee
i.e., 
\be
\label{eqpa}
\partial_\pm h(\phi) 
= {1 + \sqrt{1 + \rho}
\over \rho}\partial_\pm \rho \ \mbox{ or }\ 
{1 - \sqrt{1 + \rho}
\over \rho}\partial_\pm \rho \ .
\ee
This tells that
\be
\label{phi}
h(\phi)= \int d\rho {1 \pm 
\sqrt{1 + \rho} \over \rho} \ .
\ee
Substituting (\ref{phi}) into (\ref{eqtp2}), 
we obtain
\be
\label{ppm}
\partial_+ \partial_-\left\{
\left(1 + \rho\right)^{3 \over 2}\right\}=0
\ee
i.e.,
\be
\label{rho}
\rho=-1 + 
\left(\rho^+(x^+) + \rho^-(x^-)\right)^{2 \over 3}\ .
\ee
Here $\rho^\pm$ is an arbitrary function of 
$x^\pm = t \pm x$. 
We can straightforwardly confirm that the solutions 
(\ref{phi}) and (\ref{rho}) satisfy (\ref{req2}).
The scalar curvature is given by
\bea
\label{sR}
R&=&8\e^{-2\rho}\partial_+\partial_-\rho \nn
&=& -4{\e^{-2\left\{
-1+\left(\rho^+(x^+)+\rho^-(x^-)\right)^{2 \over 3}
\right\}} \over 
\left(\rho^+(x^+)+\rho^-(x^-)
\right)^{{4 \over 3}}}{\rho^+}'(x^+){\rho^-}'(x^-)\ .
\eea
Note that when $\rho^+(x^+)+\rho^-(x^-)=0$, 
there is a curvature singularity. 
Especially if we choose
\be
\label{Krus}
\rho^+(x^+)={r_0 \over x^+}\ ,\ \ 
\rho^-(x^-)=-{x^- \over r_0}
\ee
there are curvature singularities at $x^+x^-=r_0^2$ and 
horizon at $x^+=0$ or $x^-=0$. 
Hence we got black hole solution in the model 
under discussion.
The asymptotic flat 
regions are given by $x^+\rightarrow +\infty$ ($x^-<0$) 
or $x^-\rightarrow -\infty$ ($x^+>0$).
Therefore we can regard $x^\pm$ as corresponding to  
the Kruskal coordinates in 4 dimensions.

%%%%%%%%%%%
We now consider the Hawking radiation.
The quantum part of the energy momentum tensor 
for the generalized dilatonic supergravity is given by
\bea
\label{Tppe}
T^q_{\pm\pm}&=&{N \over 8}\left( \partial_\pm^2 \rho 
- \partial_\pm\rho \partial_\pm\rho \right) \nn
&& +{N \over 8} \left\{
\left(\partial_\pm h(\phi) \partial_\pm
h(\phi) \right)\rho
+{1 \over 2}{\partial_\pm \over \partial_\mp}
\left( \partial_\pm h(\phi) 
\partial_\mp h(\phi) 
\right)\right\} \nn
&& +{N \over 8}\left\{ 
-2 \partial_\pm \rho \partial_\pm h(\phi) 
+\partial_\pm^2 h(\phi) \right\} \nn
&&+ {N \over 64}{\partial_\pm \over \partial_\mp}
\left( h'(\phi)^2F^2 \right) + t(x^\pm) \ , \\
\label{Tpm2}
T^q_{\pm\mp}&=& -{N \over 8}\partial_+\partial_- \rho
-{N \over 16}\partial_+ h(\phi) \partial_- 
h(\phi) 
-{N \over 8}\partial_+\partial_- h(\phi) \nn
&& -{N \over 64}h'(\phi)^2F^2 
+\left({N \over 16}US 
+ {N \over 2}(-h(\phi)S^2 + h'(\phi)FS)\right)\e^{2\rho} \ .
\eea
Here we consider the bosonic background and 
put the fermionic fields to be zero.
We now investigate the case that
\be
\label{BHch}
f(\phi)=\e^{\alpha\phi}\ \  (h(\phi)=\alpha\phi)\ .
\ee
Substituting the classical black hole solution
which appeared in the original CGHS model \cite{12}
\bea
\label{sws}
\rho&=&-{1 \over 2}\ln \left(1 + {M \over \lambda}
\e^{\lambda (\sigma^--\sigma^+ )} \right) \ , \\
\phi&=&- {1 \over 2}\ln \left( {M \over \lambda}
+ \e^{\lambda(\sigma^+ - \sigma^-)} \right)  \ .
\eea
(Here $M$ is the mass of the black hole and we used asymptotic flat 
coordinates.)
and using eq.(\ref{CGHSaf}), 
we find 
\bea
\label{Tmsw}
T^q_{+-} &=& {N\lambda^2 \over 64}(4+4\alpha+\alpha^2)
{1 \over \ds^2} \nn
&& -{N\lambda^2 \over 16}(1+\alpha){1 \over \ds}
-{N\lambda^2\alpha^2 \over 64}  \ , \nn
T^q_{\pm\pm}
&=&-{N\lambda^2 \over 32} \left\{ 1 - {1 \over \ds^2}
\right\} \nn
&&-{N\lambda^2\alpha^2 \over 16}{\ln \ds - 1 \over \ds^2}
+t^\pm (\sigma^\pm) \ .
\eea
Then when $\sigma^+\rightarrow +\infty$, the energy 
momentum tensor behaves as
\bea
\label{asT}
T^q_{+-}&\rightarrow& 0
 \ ,\nn
T^q_{\pm\pm}&\rightarrow& {N\lambda^2 \over 16}
\alpha^2 + t^\pm(\sigma^\pm)\ .
\eea
In order to evaluate $t^\pm(\sigma^\pm)$, 
we impose a boundary condition that 
there is no incoming energy.
 This condition requires that $T^q_{++}$ 
should vanish at the past null infinity   
($\sigma^-\rightarrow -\infty$) and if we assume $t^-(\sigma^-)$ is 
black hole mass independent, 
$T^q_{--}$ also should vanish at 
the past horizon ($\sigma^+\rightarrow -\infty$) after taking 
$M \rightarrow 0$ limit. 
Then we find
\be
\label{t-}
t^-(\sigma^-)=-{N\lambda^2\alpha^2 \over 16}
\ee
and one obtains
\be
\label{rad}
T^q_{--}\rightarrow 0
\ee
at the future null infinity 
($\sigma^+\rightarrow +\infty$). 
Eqs.(\ref{asT}) and (\ref{rad}) might 
tell that there is no 
Hawking radiation in the dilatonic supergravity model 
under discussion when quantum back-reaction of 
matter supermultiplet in large-$N$ approach 
is taken into account. (That indicates that 
above black hole is extremal one).
This might be the result of the positive energy theorem \cite{PS}. 
If Hawking radiation is positive and mass independent, the energy 
of the system becomes unbounded below. On the other hand, the negative 
radiation cannot be accepted physically.
Of course, this result may be changed in next order of large-$N$ 
approximation or in other models of dilatonic supergravity.
From another side since we work in strong coupling regime 
it could be that new methods to study Hawking radiation should be
developed.

\section{Discussion}

In summary, we studied 2d dilatonic supergravity with dilaton coupled 
matter and dilaton supermultiplets. 
Some results of this work have been shortly reported in \cite{NOsdg}.
Trace anomaly and induced effective 
action for matter supermultiplet as well as large-$N$ effective action 
for dilatonic supergravity are calculated. Using these results one 
can estimate matter quantum corrections in the study of black holes 
and their properties like Hawking radiation. Such study is presented 
on the example of supersymmetric CGHS model which corresponds to
specific choice of generalized dilatonic couplings in the initial 
theory. Similarly, one can investigate quantum spherical collapse 
for different 4d or higher-dimensional supergravities using 2d models.

It is interesting to note that there are following directions to 
generalize our work. First of all, one can consider extended 2d 
supergravities with dilaton coupled matter. General structure 
of trace anomaly and effective action will be the same. Second, 
one can consider other types of black hole solutions in the 
model under discussion with arbitrary dilaton couplings. Unfortunately,
since such models are not exactly solvable one should usually 
apply numerical methods for study of black holes and their properties. 
Third, it could be important to discuss the well-known C-theorem 
for dilaton dependent trace anomaly. We hope to investigate some of these 
questions in near future.

\ 

\noindent
{\bf Acknoweledgments} We would like to thank R.Bousso, I. Buchbinder,
S. Hawking, S.Gates and K. Stelle for useful remarks.
This research has been supported in part by COLCIENCIAS (Colombia) and
RFBR, project No 96-02-16017 (Russia).


\begin{thebibliography}{99}
\bibitem{12} C.G. Callan, S.B. Giddings, J.A. Harvey 
and A. Strominger, {\sl Phys. Rev.} {\bf D45} (1992) 
1005.
\bibitem{RST} J.G. Russo, L. Susskind and 
L. Thorlacius, {\it Phys.Lett.} {\bf B292} (1992) 13; 
{\it Phys.Rev.} {\bf D47} (1993) 533.
\bibitem{A} S.P. de Alwis, {\it Phys.Lett.} {\bf B289} (1992) 278;
A. Bilal and C. Callan, {\it Nucl.Phys.} {\bf B394} (1993) 73;
S. Nojiri and I. Oda, {\it Phys.Lett.} {\bf B294} (1992) 317;
{\it Nucl.Phys.} {\bf B406} (1993) 499;
T. Banks, A. Dabholkar, M. Douglas and M. O`Loughlin, 
{\it Phys.Rev.} {\bf D45} (1992) 3607;
R.B. Mann, {\it Phys.Rev.} {\bf D47} (1993) 4438;
D. Louis-Martinez and G. Kunstatter, {\it Phys.Rev.} {\bf D49} 
(1994) 5227; 
T.Klobsch and T.Strobl, {\it Class.Quant.Grav.} {\bf 13} 
(1996) 965;
G. Amelino-Camelia, L. Griguolo and D. Seminara,
{\it Phys.Lett.} {\bf B371} (1996) 41.
\bibitem{BPP} S. Bose, L. Parker and Y. Peleg, 
{\it Phys.Rev.} {\bf D52} (1995) 3512;
M. Katanaev, W. Kummer and H. Liebl, {\it Phys.Rev.} {\bf D53} 
(1996) 5609.
\bibitem{Rev} T. Banks, Spring School on Supersymmetry 
and Superstrings, hep-th/9412139;
A. Strominger, Les Houches lectures on black holes, 
hep-th/9501071;
S. Giddings,  Summer School in High Energy Physics 
and Cosmology, Trieste, hep-th/9412138.
\bibitem{8} S.P. Trivedi, {\it Phys.Rev.} 
{\bf D47} (1993) 4233;
A. Strominger and S.P. Trivedi, {\it Phys.Rev.} 
{\bf D48} (1993) 2930.
\bibitem{9} G.W. Gibbons, {\it Nucl.Phys.} {\bf B207} 
(1982) 337;
G.W. Gibbons and K. Maeda, {\it Nucl.Phys.} {\bf B298} 
(1988) 741;
S.B. Giddings and A. Strominger, 
{\it Phys.Rev.Lett.} {\bf 67} (1991) 1930;
D. Garfinkle, G.T. Horowitz and A. Strominger, 
{\it Phys.Rev.} {\bf D43} (1991) 3140.
\bibitem{6} E. Elizalde, S. Naftulin and 
S.D. Odintsov, {\it Phys. Rev.} {\bf D49} 
(1994) 2852.
\bibitem{4} S. Nojiri and S.D. Odintsov, hep-th/9706009,
{\it Mod.Phys.Lett.} {\bf A12} (1997) 2083;  
hep-th/9706143, to appear in {\it Phys.Rev.} {\bf D}.
\bibitem{5} R. Bousso and S.W. Hawking, hep-th/9705236, 
{\it Phys.Rev.} {\bf D56} 7788;
S. Ichinose, hep-th/9707025; W. Kummer, H. Liebl 
and D.V. Vassilevich, hep-th/9707041, {\it Mod.Phys.Lett}
{\bf A12} (1997) 2683.
\bibitem{WB} J. Wess and J. Bagger, ``Supersymmetry 
and Supergravity'', Princeton University Press
\bibitem{GGRS} S.J. Gates, M.T. Grisaru, M. Ricek and W. Siegel, 
``Superspace or One Thousand and One Lessons in Supersymmetry'', 
Benjamin/cummings (1983) (Frontiers in Physics, 58).  
\bibitem{HUY}  K. Higashijima, T. Uematsu and Y.Z. Yu, 
{\it Phys.Lett.} {\bf 139B} (1994) 161; 
T. Uematsu, {\it Z.Phys.} {\bf C29} (1985) 143; 
T. Uematsu, {\it Z.Phys.} {\bf C32} (1986) 33.
\bibitem{NO} Shin'ichi Nojiri and Ichiro Oda, 
{\it Mod.Phys.Lett.} {\bf A8} (1993) 53.
\bibitem{MNSN} Mihoko M. Nojiri and Shin'ichi Nojiri, 
{\it Prog.Theor.Phys.} {\bf 76} (1986) 733.
\bibitem{NO2}S. Nojiri and S.D. Odintsov, 
{\sl Mod.Phys.Lett.} {\bf A12} (1997) 925.
\bibitem{PS} Y. Park and A. Strominger, 
{\it Phys.Rev.} {\bf D47} (1993) 1566.
\bibitem{NOsdg} S. Nojiri and S.D. Odintsov, hep-th/9708139, 
to appear in {\it Phys.Lett.} {\bf B}.
\end{thebibliography}
\end{document}